%% file: IEEEConf_main.tex
\documentclass[10pt, conference]{IEEEtran}

\makeatletter
\def\ps@headings{%
\def\@oddhead{\mbox{}\scriptsize\rightmark \hfil \thepage}%
\def\@evenhead{\scriptsize\thepage \hfil \leftmark\mbox{}}%
\def\@oddfoot{}%
\def\@evenfoot{}}
\makeatother
\IEEEoverridecommandlockouts
\usepackage{cite}
\usepackage{amsmath,amssymb,amsfonts}
\usepackage{algorithmic}
\usepackage{graphicx}
\usepackage{textcomp}
\usepackage{xcolor}
\usepackage{color,comment}
\usepackage{mathtools}
\usepackage[caption=false,font=footnotesize]{subfig}

\def\BibTeX{{\rm B\kern-.05em{\sc i\kern-.025em b}\kern-.08em
    T\kern-.1667em\lower.7ex\hbox{E}\kern-.125emX}}

\graphicspath{{./Camera-ready-fig/},{./fig/}}

\usepackage{amsfonts}

\newcommand{\RanSym}{T}
\DeclarePairedDelimiter\ceil{\lceil}{\rceil}

\definecolor{DarkGreen}{rgb}{0.1,0.5,0.1}
\definecolor{DarkRed}{rgb}{0.5,0.1,0.1}
\definecolor{DarkBlue}{rgb}{0.1,0.1,0.5}
\definecolor{DarkPurple}{rgb}{0.5,0.2,0.5}
\definecolor{DarkTurquoise}{rgb}{0.1,0.5,0.5}

\definecolor{beaublue}{rgb}{0.74, 0.83, 0.9}
\definecolor{coolblack}{rgb}{0.0, 0.18, 0.39}
\definecolor{apricot}{rgb}{0.98, 0.81, 0.69}
\definecolor{burntorange}{rgb}{0.8, 0.33, 0.0}
\definecolor{blue-violet}{rgb}{0.54, 0.17, 0.89}
\definecolor{byzantium}{rgb}{0.44, 0.16, 0.39}
\definecolor{brilliantrose}{rgb}{1.0, 0.33, 0.64}
\definecolor{cerisepink}{rgb}{0.93, 0.23, 0.51}
\definecolor{cobalt}{rgb}{0.0, 0.28, 0.67}
\definecolor{bostonuniversityred}{rgb}{0.8, 0.0, 0.0}

\newcommand{\off}[1]{}

\begin{document}

\title{Absolute Security \\in High-Frequency Wireless Links
}

\author{Alejandro Cohen\IEEEauthorrefmark{1}, Rafael G. L. D’Oliveira\IEEEauthorrefmark{2}, Chia-Yi Yeh\IEEEauthorrefmark{3}, Hichem Guerboukha\IEEEauthorrefmark{4}, Rabi Shrestha\IEEEauthorrefmark{4},\\ Zhaoji Fang\IEEEauthorrefmark{4}, Edward Knightly\IEEEauthorrefmark{3}, Muriel M\'{e}dard\IEEEauthorrefmark{5}, and Daniel~M.~Mittleman\IEEEauthorrefmark{4}\vspace{0.2cm}\\
\IEEEauthorrefmark{1}Faculty of Electrical and Computer Engineering, Haifa, Technion, Israel \\
\IEEEauthorrefmark{1}School of Mathematical and Statistical Sciences, Clemson University, Clemson, USA \\
\IEEEauthorrefmark{3}Department of Electrical and Computer Engineering, Rice University, Houston, SC USA \\ \IEEEauthorrefmark{4}School of Engineering, Brown University, Providence, RI USA \\
\IEEEauthorrefmark{5}Research Laboratory of Electronics, MIT, Cambridge, MA USA\vspace{-0.4cm}}

\maketitle

\input{abstract}


\input{./1_Intro}
\input{./2_AnteConf}
\input{./3_BlindRegion}
\input{./4_SecureEnc}
\input{./5_CommEff}

\input{./Note_2}
\input{./6_Experimental}

\input{conclusions}

\bibliographystyle{IEEEtran}
\bibliography{Ref}

\end{document}

%% file: abstract.tex

\begin{abstract}
Security against eavesdropping is one of the key concerns in the design of any communication system. Many common considerations of the security of a wireless communication channel rely on comparing the signal level measured by Bob (the intended receiver) to that accessible to Eve (an eavesdropper). Frameworks such as Wyner's wiretap model ensure the security of a link, in an average sense, when Bob's signal-to-noise ratio exceeds Eve's. Unfortunately, because these guarantees rely on statistical assumptions about noise, Eve can still occasionally succeed in decoding information. The goal of achieving \emph{exactly zero} probability of intercept  over an engineered region of the broadcast sector, which we term absolute security, remains elusive. Here, we describe the first architecture for a wireless link which provides absolute security. Our approach relies on the inherent properties of broadband and high-gain antennas, and is therefore ideally suited for implementation in millimeter-wave and terahertz wireless systems, where such antennas will generally be employed. We exploit spatial minima of the antenna pattern at different frequencies, the union of which defines a wide region where Eve is guaranteed to fail regardless of her computational capabilities, and regardless of the noise in the channels. Unlike conventional zero-forcing beam forming methods, we show that, for realistic assumptions about the antenna configuration and power budget, this absolute security guarantee can be achieved over most possible eavesdropper locations. Since we use relatively simple frequency-multiplexed coding, together with the underlying physics of a diffracting aperture, this idea is broadly applicable in many contexts. 
\end{abstract}

%% file: 1_Intro.tex
\section{Introduction} \label{Intro}

Concerns about wireless security date back to Marconi, when critics pointed out that if wireless signals propagate in all directions, then an adversary can also receive them \cite{marconi}. Modern wireless technologies have now begun to employ higher frequencies, in the millimeter-wave \cite{valkonen2018compact,ullah2020development,rappaport2019wireless} and terahertz ranges \cite{ma2018security,sengupta2018terahertz,ma2018invited,alexiou2020thz,standardsdoc}, which are likely to require the use of high-gain antennas to produce directional beams \cite{ma2018invited,9269933,Federici2010,sengupta2018terahertz,Peng2020}. Although this directionality inhibits eavesdropping, successful attacks are still possible since most highly directional antennas exhibit side lobe emission which sends signals in many directions. Efforts to scramble the information contained in side lobes \cite{lu20204} can offer significant improvements, but even so, an eavesdropper (Eve) will always have a non-zero probability of intercepting and decoding the transmitted message between the sender (Alice) and the intended receiver (Bob). In essence, all such security schemes rely on assumptions about noise in Eve’s measurement \cite{wyner1975wire, bloch2011physical,cohen2016wiretap,cohen2018secured}, or on her computational capabilities \cite{bernstein2008attacking,bernstein2009introduction,PQCRYPTO2015}. Despite the fact that many of these security schemes are termed in the literature as exhibiting perfect security \cite{bloch2011physical,liang2009information,zhou2013physical}, it is clearly more favorable if Eve has \emph{zero} probability of intercepting the message from Alice to Bob, regardless of assumptions. 

\begin{figure*}[ht]
  \centering \includegraphics[trim={1.5cm 2.3cm 1.6cm 2.0cm},clip, width=0.85\textwidth]{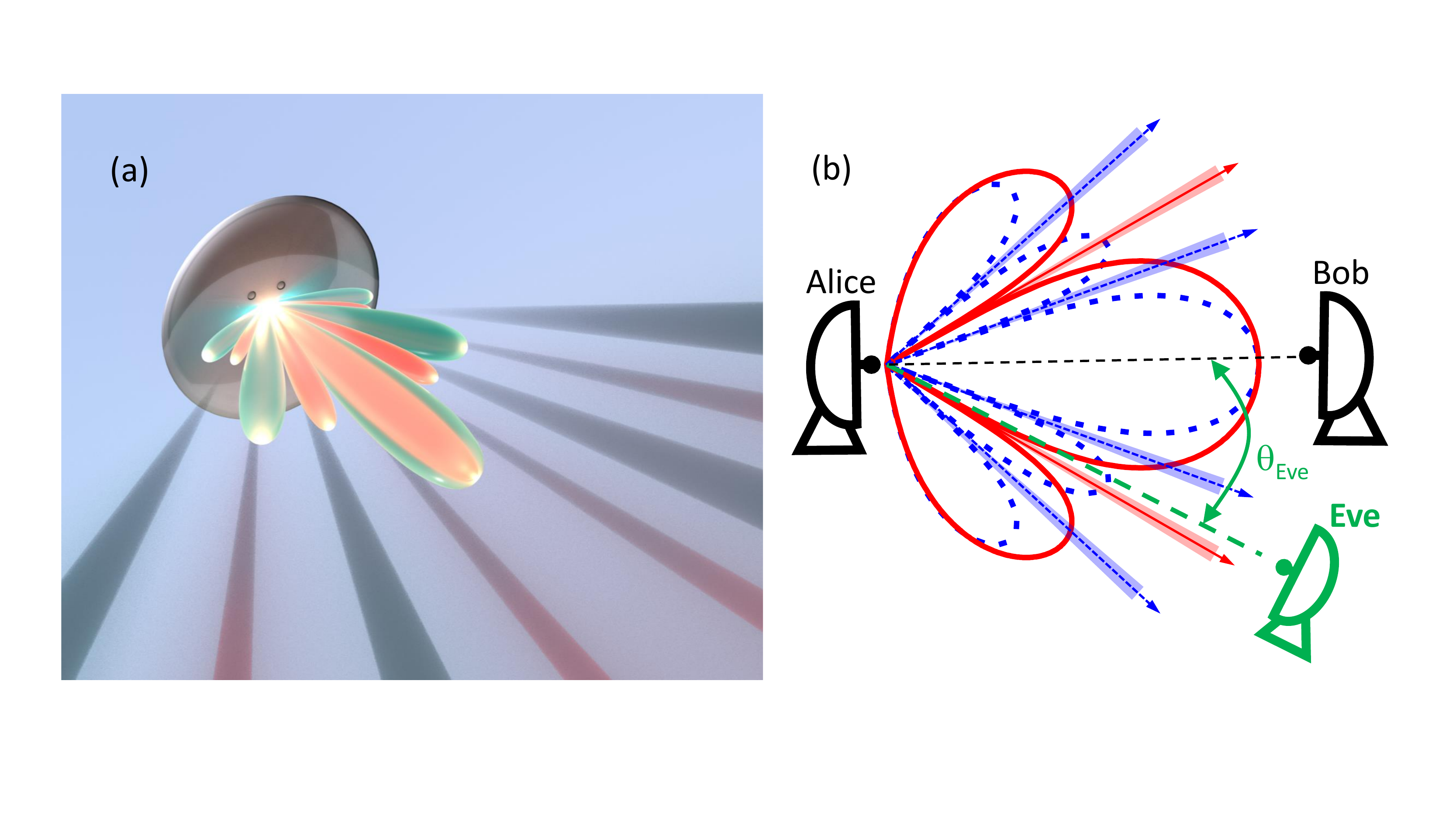}
  \caption{(a) An illustration of radiation patterns from a parabolic dish at two different frequencies, showing the main lobe and side lobes. The minima of each pattern define angular regions where signals cannot be detected at that frequency. With many subchannel frequencies, the union of these minima creates a blind region covering most of the possible locations for an eavesdropper. (b) Schematic diagram illustrating these minima, as Alice (the transmitter) broadcasts to Bob (the intended receiver), attempting to thwart Eve (an eavesdropper) located at angle $\theta_{Eve}$.
}
\label{fig:Artwork}
\end{figure*}

In this paper, we describe a new approach to realize what we term \emph{absolute} \off{post-quantum }security, which we define\off{ \emph{absolute security}, of any type, } as security\off{ for that type } \emph{with probability one}. Such a notion contrasts with common probabilistic security typically used in physical layer security discussions, that holds merely in an average sense over noise realizations, thus permitting some transmissions to be insecure. In contrast, absolute security holds with probability one for any realization of the noise, even for the putative case, most favorable to Eve, in which her measurement is noiseless.

Our approach to achieve absolute security relies on both the inherent properties of Alice's antenna and on an associated secure coding scheme\off{, which can be implemented without requiring significant computational overhead or incurring appreciable data rate penalty}.
Many directional antennas, when driven over an ultrawide bandwidth, result in frequency-dependent minima (see Fig.~\ref{fig:Artwork}). Since any receiver has a minimum detectable signal threshold, radiation minima create regions in space where Eve cannot even detect the signal, regardless of the noise realization.
This allow us to leverage recent developments in secure communications to thwart Eve as long as some frequencies are ``blind'' for her.
This approach enables Alice and Bob to establish a secure wireless link that cannot be broken by any adversary located in an engineered region of the broadcast space, even if she possesses arbitrarily powerful computational capabilities, even a quantum computer \cite{shor1999polynomial,hallgren2005fast,hallgren2007polynomial,schmidt2005polynomial,cohen2021network}.

Our discussion mostly focuses on the situation in which the eavesdropper is located within the engineered region of the broadcast sector that enables absolute security (the so-called ``blind region'', defined in Section~\ref{BlindRegion}). We also describe a small modification to our method which allows us to provide absolute security even against eavesdroppers who are \emph{not} in this blind region.
Our method breaks the conventional paradigm for secure communications in which one \off{typically}faces a trade-off between data transmission rate and the degree of security: in our approach, increasing the transmission bandwidth (and therefore the achievable data rate) can \emph{simultaneously} offer improved security. This method is therefore particularly well suited for future generations of wireless technology, which will exploit ultra-wideband channels in the millimeter-wave and terahertz regions of the spectrum \cite{sengupta2018terahertz}.

To demonstrate our proposed method, we perform model-driven analysis for multiple antennas suitable for millimeter-wave and terahertz band, as well as experimental measurement with over-the-air data transmissions. With model-driven analysis, we show how the blind region increases with a larger bandwidth, when the antenna features frequency-dependent minima, including phased arrays, parabolic dishes, and leaky-wave antennas. We also show that not all antennas are suitable for our proposed method. Horn antennas, for example, do not exhibit pronounced minima, and thus increasing bandwidth does not enlarge the blind region. However, in the experiment, we show that the horn antenna can still be used for our method. By placing a beam block in front of the horn antenna, we create diffraction pattern and pronounced frequency-dependent minima. With three widely spaced frequencies (100, 200, and 400 GHz), we demonstrate a substantial blind region where Eve fails to detect at least one of the three modulated data streams and thus achieve absolute security.

Lastly, we contrast our security scheme with a conventional method known as zero-forcing, in which a phased array is engineered to create a minima in the radiation pattern at a specific location in order to thwart an eavesdropper at that location. Our approach is quite distinct from this legacy approach, for several reasons. As detailed below, the blind region in our method is the union of minima over all frequency channels. Thus, we do not need to know the precise location of the eavesdropper, only whether she is located in this blind region (which can realistically encompass a large fraction of the full angular range). Moreover, unlike the case of zero-forcing, if Eve fails to measure just one of the frequency channels in our approach, she is unable to decode any of them.

%% file: 2_AnteConf.tex
\section{Absolute Security} \label{AbsoluteSec}
\subsection{Antenna Configuration} \label{AnteConf}

For many antennas, the far-field radiation pattern exhibits minima in specific directions, which depend on the details of the antenna geometry and its excitation mechanism, as well as on the frequency of the radiation \cite{balanis2016antenna}. 
For example, two commonly employed antennas in high-frequency wireless links, a linear phased array \cite{802.11ay,5g-will-work} and a center-fed parabolic dish \cite{9269933}, both exhibit pronounced minima at various angles, which shift with transmission frequency (see Section~\ref{Note_1}).
Under the assumption (discussed further below) that Eve must avoid \emph{all} of these minima, a transmission with multiple frequency bands creates a significant excluded region for Eve. To quantify this, we consider a transmission that uses a bandwidth $B$ from $f_L$ to $f_H$, centered on $f_C = (f_L+f_H)/2$,  sliced uniformly into $q$ frequency channels, each with bandwidth  $w = (f_H-f_L)/q$.  At location $(r,\theta)$, the received intensity $S$ (in W/$m^2$) in the $i$-th frequency channel $[f_i - \frac{w}{2}, f_i+\frac{w}{2}]$ can then be represented as
\begin{equation}
S(f_i) \propto \int_{f_i-\frac{w}{2}} ^{f_i+\frac{w}{2}} P_T(f) \cdot \gamma(r,f) \cdot G(f,\theta) df \label{eq: signal} 
\end{equation}
where $P_T(f)$ is the transmit power spectrum (in W/Hz) employed by Alice, $\gamma(r,f)$ is the distance- and frequency-dependent channel gain from the transmitter to the receiver and $G(f,\theta)$ is the antenna radiation pattern. For simplicity, we consider only one emission plane (H plane), although our results can readily be generalized to three dimensions. 

%% file: 3_BlindRegion.tex
\subsection{Defining the Blind Region} \label{BlindRegion}

For any receiver, there exists a minimum detectable signal threshold $\delta > 0$ (intensity per unit bandwidth), below which the receiver cannot detect a transmission. This threshold may depend on the receiver sensitivity, the receive antenna gain, the environmental noise floor, and the quantization of digital processing (see Section~\ref{Note_2}). The existence of this non-zero threshold $\delta$ implies that there are \textbf{blind regions} where, with probability one, Eve cannot detect the transmission. We define the blind region ($\Omega$) for a transmission band $[f_L, f_H]$ as the set of locations $(r_{Eve},\theta_{Eve})$ where Eve is unable to detect signals in \emph{at least one} of the $q$ frequency channels. Specifically, we first define the blind region for the $i$-th frequency channel as the set of locations for which the signal intensity is below the detection threshold:
\begin{equation}
\mathcal{Z}(f_i) = \{ (r_{Eve}, \theta_{Eve}) | S(f_i)<\delta \cdot w  \}. \label{eq: single frequency blind region} 
\end{equation}
The blind region for the total transmission band is then the union of blind regions for each subchannel:
\begin{equation}
\Omega = \bigcup_{i=1}^{q} \mathcal{Z}(f_i),  \label{eq: band blind region}
\end{equation}
For each location in the blind region $\Omega$, the number of missing frequency channels can vary from one up to all $q$ of them. We therefore also define $\Gamma$ as the number of subchannels for which $S(f_i)<\delta \cdot w$. Each possible location for Eve can therefore be characterized as either non-blind ($\Gamma=0$) or $\Gamma$-blind ($1 \leq \Gamma \leq q$).

As the number of subchannels $q$ increases, Alice's broadcast includes more signals at distinct frequencies with unique radiation patterns, each exhibiting minima in distinct directions. Thus, the percentage of angular locations $\theta_{Eve}$ that are within the blind region also increases.

We emphasize that the blind region defined here is not just a function of the antenna and broadcast frequencies.  It also depends on the properties of Eve's receiver, through the parameter $\delta$ defined above. As a result, different assumptions about Eve's receiver capabilities will result in somewhat different blind regions. However, even in the hypothetical best case (for Eve) that her receiver is quantum-noise limited, her ability to detect Alice's broadcast is still limited by the thermal noise of the environment which she is observing (see Section~\ref{Note_2}). Of course, it is possible to detect signals that are well below the thermal background; this is commonly achieved, for example in astrophysical observations, by severely restricting the spectral bandwidth of the detection and/or extended signal averaging. However, Eve cannot employ these strategies if she wishes to decode a broadband high-data-rate transmission. Thus, the value of $\delta$ cannot be infinitesimal, regardless of how Eve detects signals. An important consequence of this conclusion is that we \emph{need not require} that Eve's location must precisely coincide with the (mathematically infinitesimal) angular position of a minimum in an antenna radiation pattern; she only needs to be \emph{close enough} to a minimum such that her received signal is small. 

This consideration emphasizes the clear distinction between our proposal and the idea of extending conventional narrowband beam forming methods based on zero forcing to a broadband context.\cite{cho2020zero, Strobe} With zero-forcing, one can engineer an antenna (e.g., the signals applied to each element of a phased array) to force the broadcast wave amplitude to zero in a given direction at a given frequency. This would make it impossible for Eve to detect signals at that frequency, if she is located in that direction. But she would still be able to detect signals at other frequencies, since the zero is enforced in her direction only for one particular frequency. By contrast, with our method, Eve would fail to decode any of the frequency channels, not merely the one whose antenna pattern is forced to be zero at her location. Indeed, our approach does not require knowledge of Eve's location. Since the blind region defined by Eq. \eqref{eq: band blind region} is the \emph{union} of minima over all frequency bands, it can quite realistically occupy a significant fraction of the total angular space. The approach described here scales favorably with increasing transmission bandwidth, while the exact opposite is true for security schemes based on zero-forcing. It is also worth noting that zero-forcing only works for phased arrays; meanwhile, our approach has the advantage of working well for many antenna configurations, including for instance a conventional parabolic dish antenna (see Fig. \ref{fig:SecExample}(b)), where zero-forcing techniques obviously cannot be applied.

It is the coordinated use of, on the one hand, the union of blind regions $\Omega$ from frequency-dependent radiation patterns and, on the other hand, a secure coding scheme, that constitutes the core of our method's novelty. Unlike legacy methods relying on design of minima regions for security \cite{zhou2013physical}, the particular subset of frequencies that Eve can detect in any of the blind regions is irrelevant for our approach. This lack of dependence on the subset of detectable channels greatly expands the notion and, hence, footprint of the blind regions relative to traditional beam forming methods. 

%% file: 4_SecureEnc.tex
\subsection{Secure Encoding} \label{SecureEnc}

In this section, we consider the first encoding scheme, which we denote as Scheme 1, assumes that Eve is within the blind region. We illustrate the ideas by using a simplified situation in which Alice wants to communicate securely with Bob using only $q=3$ subchannels, at frequencies $f_1$, $f_2$, and $f_3$. The general idea can readily be scaled to a larger number of subchannels from known constructions in the literature \cite{cohen2018secure,silva2009universal,silva2011universal}. In the encoding scheme considered here, we assume that Eve is within the blind region. Our scheme operates symbolwise, so Alice must map her message into blocks, and then map each block into a symbol selected from a finite field of dimension greater than $2^q$ \cite{cohen2018secure}. For ease of exposition, we consider here a prime field. 
The construction can be easily generalized to operation over extensions of the binary field.
Because our simplified illustration employs $q=3$ subchannels, our illustrative example employs the finite field $\mathbb{F}_{11}$ \cite{silva2009universal,silva2011universal}. Alice first partitions her message (strings of bits) into blocks of length $\ceil{\log_2(11)}$, and then maps each block to a symbol of $\mathbb{F}_{11}$. To transmit a single message symbol $M \in \mathbb{F}_{11}$ securely to Bob, Alice first generates two symbols $\RanSym_1,\RanSym_2 \in \mathbb{F}_{11}$ uniformly at random.  Alice then generates three encoded symbols $X_1,X_2,X_3$ $\in \mathbb{F}_{11}$ using her message $M$ and the two random symbols $\RanSym_1$ and $\RanSym_2$, given by
\begin{equation} \label{eq: example coding}
\begin{split}
    X_1 &= M + \RanSym_1 + \RanSym_2, \\
    X_2 &= M + 2 \RanSym_1 + 4 \RanSym_2,  \\
    X_3 &= M + 3 \RanSym_1 + 9 \RanSym_2. 
\end{split}
\end{equation}
Each encoded symbol $X_i$ is transmitted to Bob via the frequency band $f_i$. Since Bob is not in the blind region (i.e., his location has $\Gamma=0$), he receives the three encoded symbols and is able to decode the message symbol $M$ by means of a simple linear transform which inverts Eq. \eqref{eq: example coding}:
\begin{equation} \label{eq: transform}
     \begin{pmatrix}
    M \\ \RanSym_1 \\ \RanSym_2
    \end{pmatrix}
    =
    \begin{pmatrix}
    3 & 8 & 1 \\
    3 & 4 & 4 \\
    6 & 10 & 6
    \end{pmatrix}
    \begin{pmatrix}
    X_1 \\ X_2 \\ X_3
    \end{pmatrix}.
\end{equation}
However, since Eve is in the blind region, she can observe at most two encoded symbols from the set $\{X_1,X_2,X_3\}$ with probability one. We can show that, regardless of which two {encoded symbols} Eve detects, she cannot determine $M$. For instance, if Eve receives $X_1$ and $X_2$, then the mutual information between her observations and the message symbol $M$ can be computed from the entropy as:
\begin{equation} \label{eq: example security}
\begin{split}
    I(M;X_1,X_2) &= H(X_1,X_2) - H(X_1,X_2 | M) \\
    &= H(X_1,X_2) - H(\RanSym_1+\RanSym_2,2\RanSym_1+4\RanSym_2) \\
    &= H(X_1,X_2) - 2 \log(|\mathbb{F}_{11}|) \\
    &\leq H(X_1) + H(X_2) - 2 \log(11)   \\
    &= 0. 
\end{split}
\end{equation}
This result follows directly from the definition of mutual information, and the fact that, conditioned on the messages, the only uncertainty about $X_1$ and $X_2$ is in the random variables $T_1+T_2$ and $2T_1+4T_2$, which are independent and uniform. Thus, because there is zero mutual information between Eve's observation and Alice's message, Eve learns nothing about $M$; absolute security is guaranteed.

%% file: 5_CommEff.tex
\subsection{Increasing the Secure Communication Efficiency} \label{CommEff}

We can define the secure communication efficiency in terms of the length of Alice's message. This efficiency $\eta$ is the ratio between the size of the message and the number of bits needed to transmit it. Ideally, one would like this rate to be as close to $\eta=1$ as possible. Generally, in previously proposed security schemes, this is not possible owing to the need to add redundancy to the transmission in order to guarantee security in the communication \cite{wyner1975wire,ozarow1984wire,bloch2011physical}. In the security scheme described in Section~\ref{SecureEnc}, by noting that Alice must send $q=3$ encoded  symbols to transmit the original message symbol, we see that the secure communication efficiency\off{ of Scheme 1} is $\eta = \frac{1}{3}$. In general, the efficiency scales inversely with the number of frequency channels, $\eta \propto \frac{1}{q}$

It is easy to address this issue of the less-than-ideal efficiency of our approach, by making a small modification to the method, which we term Scheme 2: Alice can replace the $q-1$ (in our example, two) random symbols with additional messages, $M_2$ and $M_3$, and then perform the same encoding as in Eq. \eqref{eq: example coding}, with the random symbols replaced by the additional messages. Alice can thus obtain an optimum secure communication efficiency of $\eta=1$, regardless of the number of channels.
That is, Alice replaces the random symbols, $T_1$ and $T_2$, with message symbols $M_2$ and $M_3$, and then transmits the three encoded symbols $X_1,X_2,X_3$ as in Eq.~\eqref{eq: example coding},\off{ with $M,T_1,T_2$ replaced by $M_1,M_2,M_3$,} i.e., 
\begin{equation*} \label{eq: example individual coding}
\begin{split}
    X_1 &= M_1 + M_2 + M_3, \\
    X_2 &= M_1 + 2 M_2 + 4 M_3,  \\
    X_3 &= M_1 + 3 M_2 + 9 M_3. 
\end{split}
\end{equation*}


As before, Bob can decode all three message symbols through a linear transform; but, the secure communication efficiency issue is now solved, since now $q=3$ encoded symbols are sent in order to retrieve $q=3$ message symbols, i.e., $\eta=1$. This scheme guarantees zero mutual information with any subset of message symbols, yet may potentially allow Eve to obtain information about linear combinations of the message symbols \cite{cohen2018secure}. In order to implement this approach, Alice must ensure that the message symbols $M_1,M_2,M_3$ are uniformly distributed. The reason for this, intuitively, is that the message symbols themselves are performing the role of the random symbols $T_1$ and $T_2$. We note that there are known techniques described in the literature \cite{hayashi2016secure} which can be used to enforce this uniformity condition, so this requirement is not a significant impediment. Thus, although $I(M_1,M_2,M_3;X_1,X_2)$ may not be zero, it is nevertheless possible for Alice to guarantee that the mutual information between any individual message and any two transmitted symbols is zero. That is, for any distinct $i,j,\ell \in \{1,2,3\}$, it follows that
\begin{equation*} 
\begin{split}
    I(M_i;X_1,X_2) &= H(X_1,X_2) - H(X_1,X_2 | M_i) \\
    &= H(X_1,X_2) - H(M_j+M_\ell,2M_j+4M_\ell) \\
    &= H(X_1,X_2) - 2 \log(|\mathbb{F}_{11}|) \\
    &\leq H(X_1) + H(X_2) - 2 \log(11)  \\
    &= 0. 
\end{split}
\end{equation*}
We stress that the information that Eve \emph{can} obtain in this situation (which involves only linear combinations of Alice's messages) is largely trivial, and cannot in general be used to decode or decipher any meaning.



A key and, to our knowledge, unique advantage of our method is that it provides improved security as the bandwidth of the transmission increases. Indeed, as $q$ increases, Alice is afforded more bandwidth which, because of the $\eta=1$ communication efficiency, increases the data rate in her link with Bob while simultaneously improving the security by expanding the size of the blind region $\Omega$. This simultaneous improvement in security and data rate has never previously been realized in wireless systems.

%% file: Note_2.tex
\subsection{Effect of Minimum Detectable Signal Threshold} \label{Note_2}

As noted, a key assumption of our approach is that a detector has a non-zero threshold $\delta$ for minimum detectable signal. This assumption is valid for any RF receiver, other than perhaps those which operate near the single-photon detection limit \cite{komiyama2010single}. For the purposes of our illustrative calculations, we can consider a conservative threshold based on thermal radiation. When staring at a room temperature (300K) blackbody, an area of 1 $\text{cm}^2$ intercepts 0.29 nW of power within a 1-GHz-wide frequency band from 100 GHz to 101 GHz, or 2.55 nW from 300 GHz to 301 GHz. In fact, most receivers employed in RF communication systems do not even approach this sensitivity (and this becomes increasingly true as the frequency increases into the millimeter-wave and terahertz regimes, where detectors are typically much less sensitive), so these values are something of a worst-case scenario.

For purposes of computing the channel capacity to Bob\off{ as in Fig. \ref{fig:TotalRate}} and to further illustrate its role in a communication system,
let us now consider the effect of $\delta$ on the channel between Alice and Bob. 
Bob must be able detect the minimum difference in fluence, $\delta$, between any two symbols. It must be that any received symbol has, by the process of detection, a minimum detection uncertainty of energy $\delta$ Joules per meter squared, since two symbols with fluence difference less than $\delta$ could not be distinguished from each other. The effect of this uncertainty is to limit Bob's throughput. We note that, in our considerations of Eve's capabilities, we place no such limit, in order to consider a very powerful eavesdropper.

Let us now consider the power per meter squared, $\sigma^2$, corresponding to this detection uncertainty. Given that we require a minimum fluence of $\delta$ to detect a signal with intensity $\sigma^2$, we require a minimum time of observation $\tau$  such that $\tau \sigma^2 \geq \delta$ in order to detect the detection uncertainty inherent to a symbol. Thus $\frac{1}{\tau}$ is the fastest sampling rate for symbols.

Let the signal intensity be denoted by $P$. The Shannon capacity, if we have only the detection uncertainty, is $\frac{1}{2 \tau} \ln \left( 1 + \frac{P}{\sigma^2} \right) $, assuming the pessimum uncertainty distribution, which is Gaussian \cite[Chapter 9]{cover2012elements}. Note that we should assume such a pessimum  distribution, since we have no guarantees on its form, only on its fluence. We may rewrite this capacity as  $\frac{1}{2 \tau} \ln \left( 1 + \frac{P}{\delta / \tau} \right) $. This expression increases as $\tau$ decreases. By Nyquist, $\tau \leq 1/w$, so we obtain a capacity of $\frac{w}{2} \ln \left( 1 + \frac{P}{\delta \cdot w} \right) $ as the maximum rate available when only the reception uncertainty is taken into account. 
If we have additional noise of the conventional form, that noise will further reduce capacity.

%% file: 6_Experimental.tex
\section{Absolute Security Evaluation} \label{Evaluation}

In this section, we evaluate the absolute security approach we propose in Sec. \ref{AbsoluteSec} using model-driven analysis and experimental demonstration.


\subsection{Model-Driven Analysis}

Since our method leverages antenna's frequency-dependent minima and coding to create blind regions, we examine the security performance when different types of antenna are employed.
The first set of antennas features a fixed main lobe direction and frequency-dependent minima. The selected antennas in this category include a linear phased array and a parabolic dish.
Next, we examine antennas whose main lobe direction shifts very strongly with frequency (in addition to frequency-dependent minima), with the example being a leaky-wave antenna.
Further, we show that not all antennas are suitable for employing the proposed absolute security approach, especially for antennas without pronounced minima, such as the horn antenna.

\subsubsection{Phased Array and Parabolic Dish} \label{Note_1}


In the subsection, we consider two specific antenna geometries to provide concrete illustrations of the ideas that underlie our security protocol.
One of these is a 16-element linear phased array, in which each element is a vertically polarized point dipole emitter, and the elements are spaced along a horizontal line by half of the center wavelength ($\lambda=1.5$mm in our simulations). The other is a parabolic dish antenna, with a diameter of 16mm and a focal length of 10mm, emitting vertically polarized radiation with a directional gain of 30.5 dBi at a frequency of 200 GHz. The phased array configuration is representative of steerable antennas that are commonly employed in today's millimeter-wave Wi-Fi and 5G standards, while the parabolic dish has often been employed in backhaul and other fixed broadband applications. In both cases, these antenna configurations scale naturally into the millimeter-wave and terahertz range, and have been employed for such high-frequency transmissions.

\begin{figure}[ht]
  \centering \includegraphics[trim={0cm 0cm 0.5cm 0cm},clip, width=0.4\textwidth]{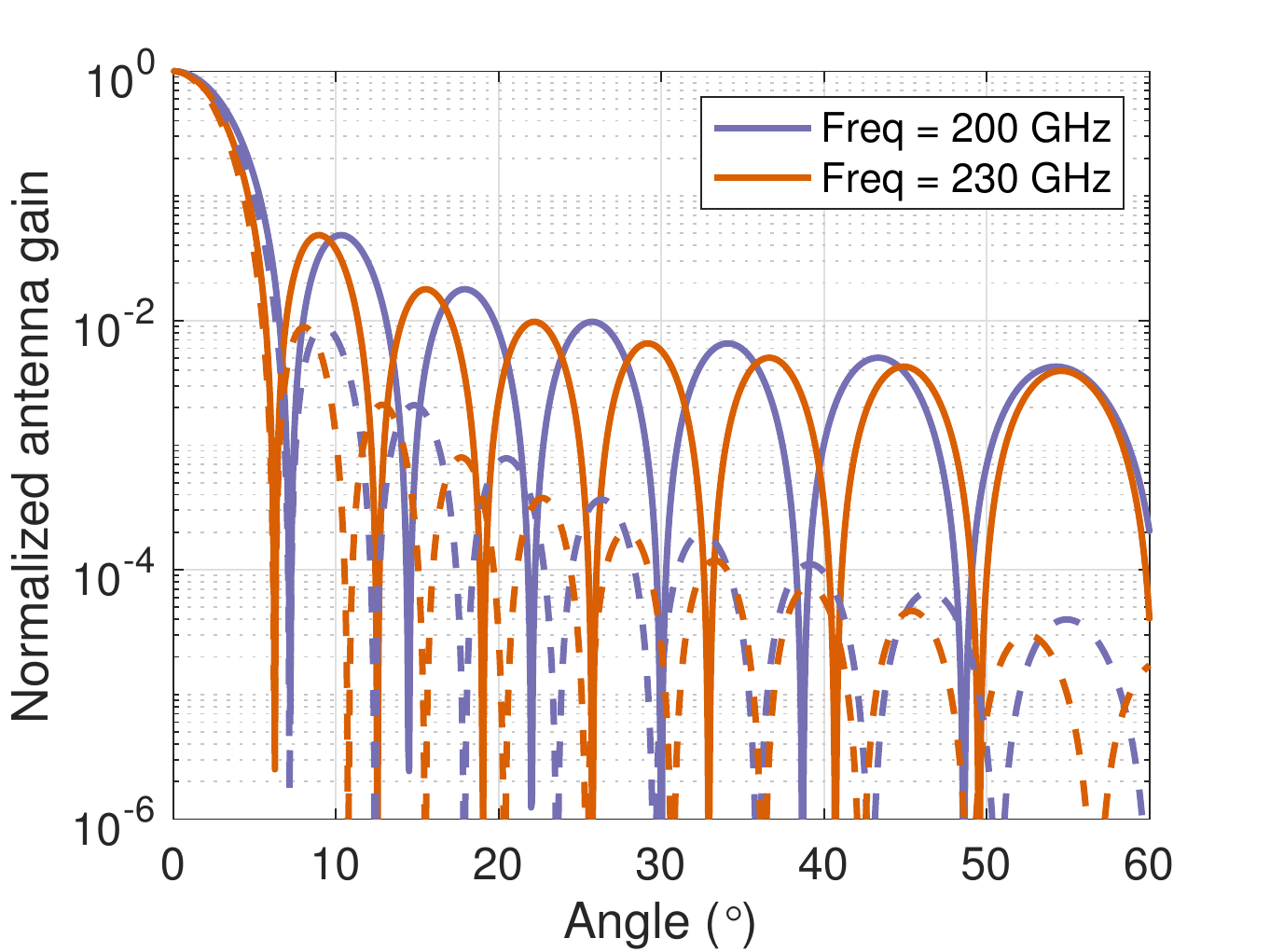}
  \caption{Radiation patterns illustrating how the pronounced minima shift with frequency (solid: phased array, dashed: parabolic dish). 
  }
\label{fig:RadPattn_Para_PA}
\end{figure}

Although radiation patterns are of course three-dimensional, for simplicity we illustrate the essential idea of our approach by only considering a two-dimensional slice (the horizontal plane which is orthogonal to the polarization axis, the H-plane), for simplicity. Fig. \ref{fig:RadPattn_Para_PA} shows the radiation patterns of the two example antennas, at two different frequencies. In this figure, we observe that, even if Alice uses only the few frequency bands shown in these illustrations, many of Eve's possible locations are ruled out by the fact that she must avoid all of the minima of every frequency in Alice's transmission.

Using the phased array and the parabolic dish as described, we calculate the absolute secure angles according to Eq. \eqref{eq: band blind region} for a transmission with
a center frequency of $f_c=200$ GHz, a subchannel bandwidth of $w=1$ GHz, and for several values of the parameter $P_{AB}$ which describes Alice's transmit power to Bob.
In particular, Alice's transmit power is parameterized by the intensity received by Bob, normalized to the detection threshold discussed above, $P_{AB}=S_{Bob}/(\delta \cdot w)$.
For this calculation, Eve is at the same distance from Alice as Bob, and Alice adjusts her transmit power so that Bob receives a fixed intensity level $S_{Bob}$ at all frequencies from $f_L$ to $f_H$.


\begin{figure}[ht]
  \centering
  \subfloat[] {\includegraphics[trim={0cm 0cm 0.5cm 0cm},clip, width=0.4\textwidth]{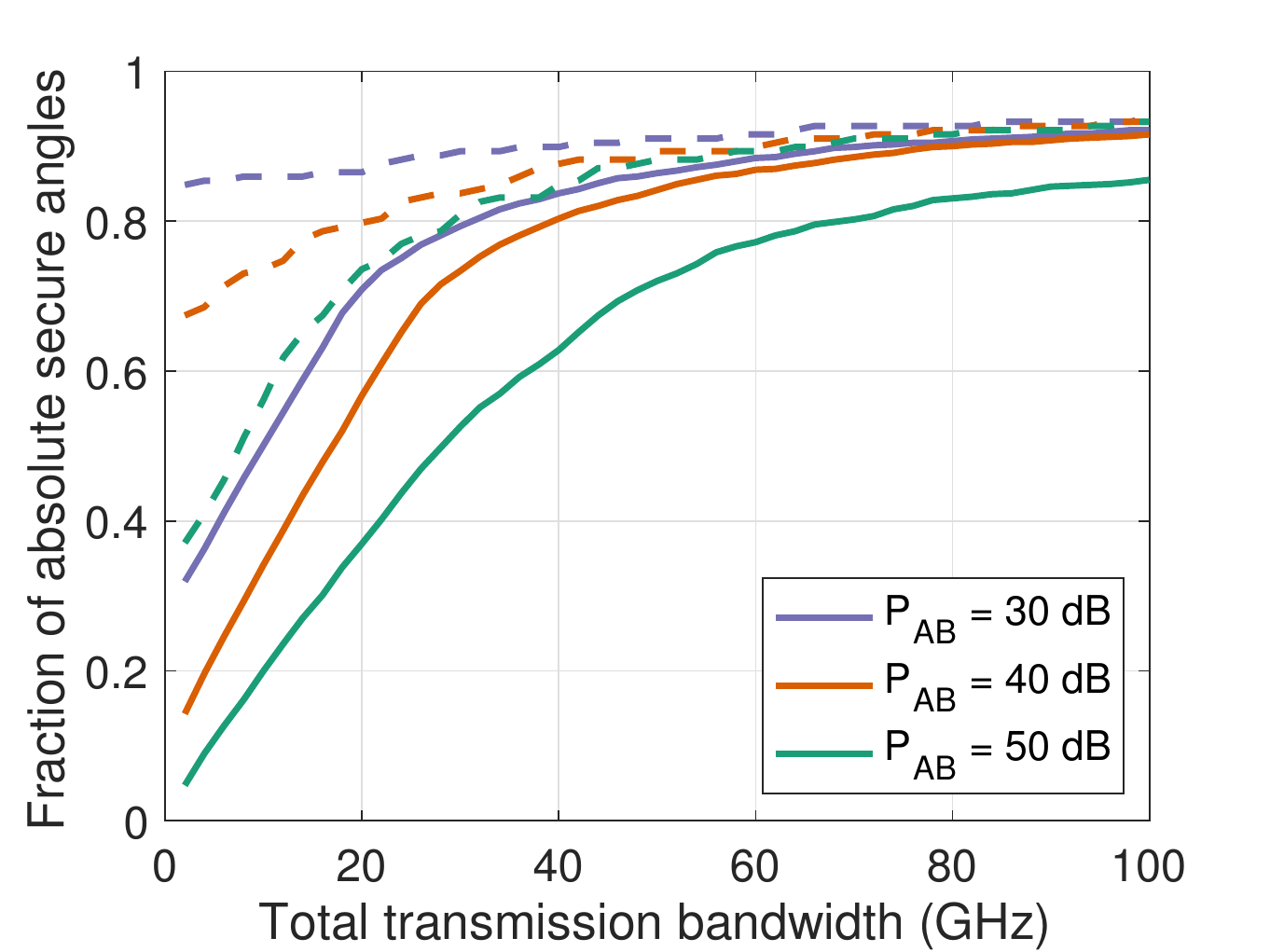}\label{fig:SecAng_TotalBW}}
  \quad
  \subfloat[] {\includegraphics[trim={0cm 0cm 0.5cm 0cm},clip, width=0.4\textwidth]{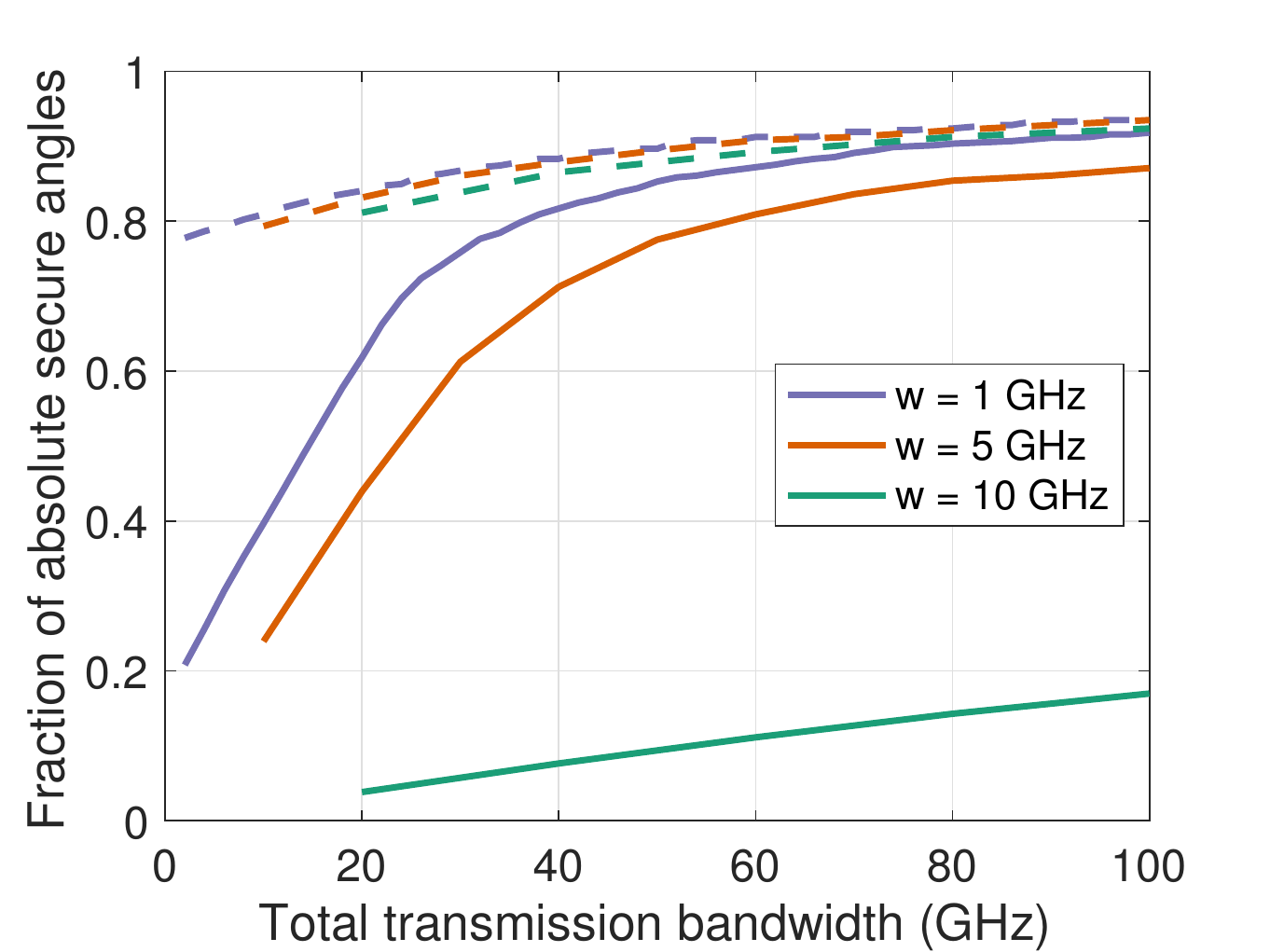}\label{fig:Channelization}}
  \caption{Size of the blind region increases with bandwidth (solid: phased array, dashed: parabolic dish). (a)
  For several values of Alice's transmit power parameterized by $P_{AB}$.
  (b)
  For different values of subchannel bandwidth $w$.}
\label{fig:SecExample}
\end{figure}

Fig.~\ref{fig:SecAng_TotalBW} illustrates the size of the blind region increase as a function of total bandwidth $B$, assuming Alice transmits to Bob using the antenna main lobe.
For an increasing transmission bandwidth, as long as Eve is outside of the main antenna lobe (where Bob is located), she is increasingly likely to be within a blind region, i.e., at least one frequency channel is below her detection threshold ($\Gamma > 0$).
In Fig.~\ref{fig:SecAng_TotalBW}, the limiting value at large bandwidth is determined by the angular width of the main lobe of the antenna pattern, where Bob is located (and which, by definition, is never within the blind region).

The width of the subchannels also impacts the size of the blind region for a given bandwidth.
Using the same setup as in Fig.~\ref{fig:SecAng_TotalBW} with a fixed transmit power parameterized by $P_{AB}=$ 35 dB, Fig.~\ref{fig:Channelization} illustrates the blind region for different subchannel bandwidths $w$.
From Fig.~\ref{fig:Channelization}, we observe that when the width of the subchannel is larger, it is harder to guarantee that the signal intensity across the subchannel is below the detection threshold, so the blind region is smaller.

\subsubsection{Angularly Dispersive Antennas}

\begin{figure}[ht]
  \centering
  \subfloat[] {\includegraphics[trim={0cm 0cm 0.5cm 0cm},clip, width=0.4\textwidth]{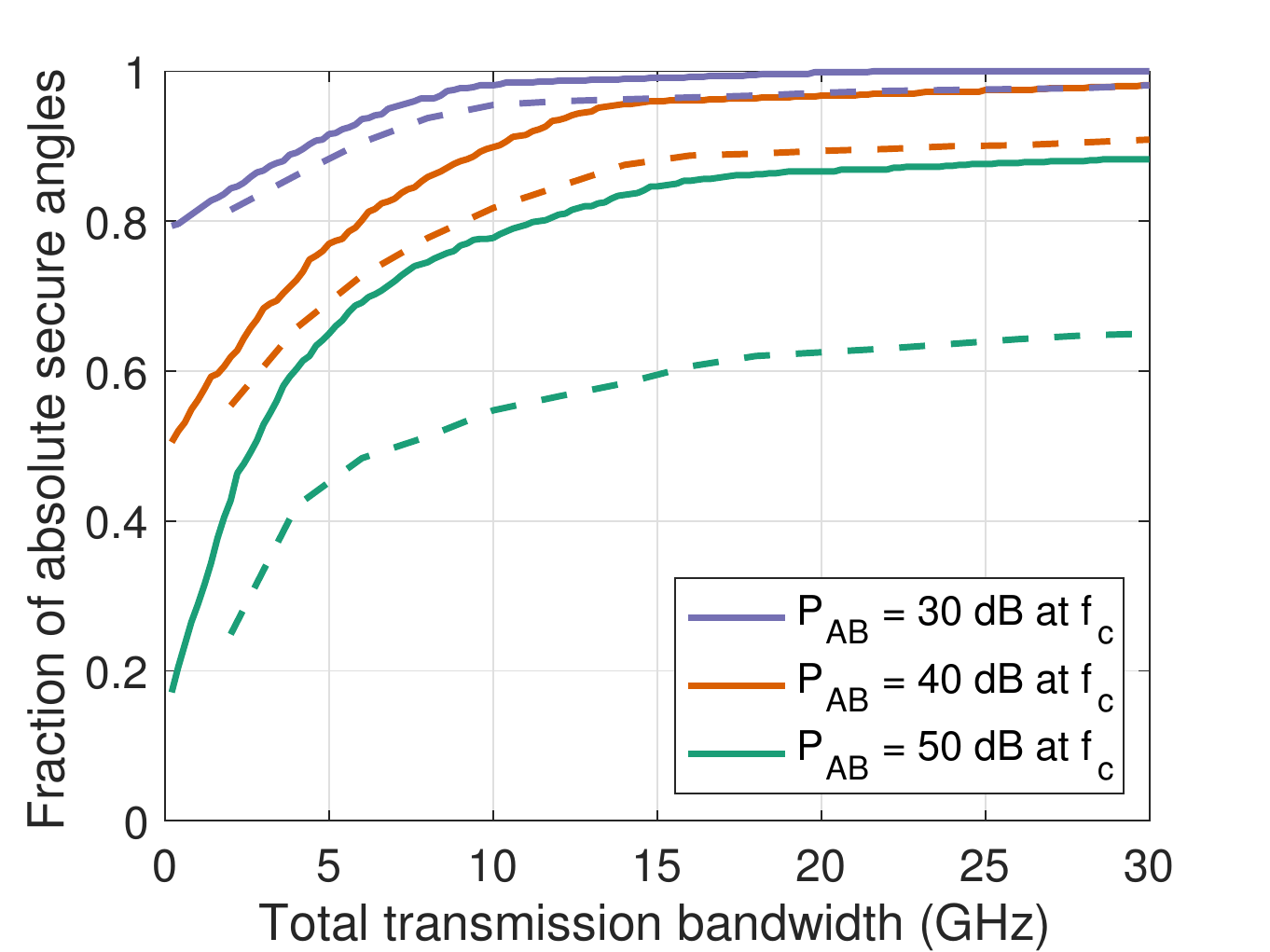}\label{fig:SecAng_TotalBW_LWA}}
  \quad
  \subfloat[] {\includegraphics[trim={0cm 0cm 0.5cm 0cm},clip, width=0.4\textwidth]{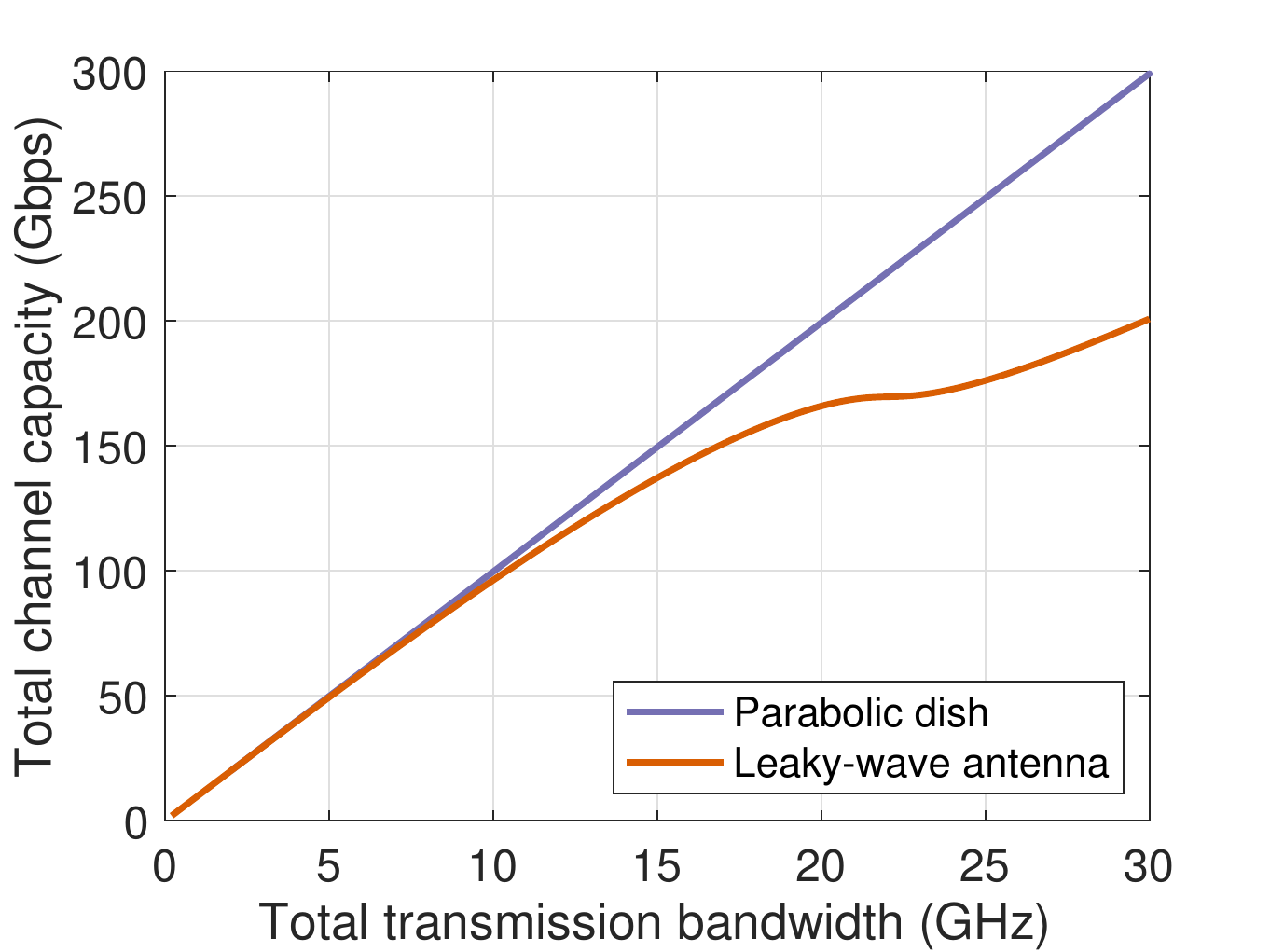}\label{fig:TotalRate}}
  \caption{\textbf{A leaky-wave antenna with strong angular dispersion.} (a) The fraction of the angular range which is within the blind region ($\Gamma > 0$), and thus offers absolute security, as a function of bandwidth for several values transmit power parameterized by $P_{AB}$ (solid: subchannel bandwidth $w=0.1$ GHz, dashed: $w=1$ GHz).
  (b) The scaling of total capacity with increasing transmission band comparing a non-angularly dispersive antenna (parabolic dish) and an angularly dispersive link (leaky-wave antenna). $P_{AB}=30$ dB is considered in both cases. Here we assume Bob and Eve have the same detection threshold and an equal antenna aperture.}
\label{fig:LWA}
\end{figure}


We address the possibility of implementing the same security scheme using a different class of antenna structure, in which the main lobe of the broadcast shifts very strongly with frequency. A prototype of such an antenna is a leaky-wave waveguide, which exhibits very strong angular dispersion \cite{caloz}.

We employ a parallel-plate leaky-wave antenna with a plate separation of 1 mm and an attenuation constant of 1. 
Bob is located at $30^\circ$ and the maximum radiation frequency at this angle, 300 GHz, is the center frequency for the transmission.
For the illustrative calculation, we employ two values of subchannel bandwidth, $w=\{0.1, 1\}$ GHz.
As above, we assume that Eve is at the same distance from Alice as Bob. Notice that $P_{AB}$ varies across the transmission band due to the dispersive link when Alice employs a uniform transmit power. Thus, we use the value of $P_{AB}$ corresponding to the center frequency to characterize the transmit power.

With angular dispersion, the available bandwidth for transmissions between Alice and Bob is restricted, since widely differing frequencies propagate in very different directions. As a result, although the blind region still increases with the transmission bandwidth (Fig. \ref{fig:SecAng_TotalBW_LWA}) and Scheme 2 can still be employed in the blind region, there is a limit to the improvement in data rate (see Fig. \ref{fig:TotalRate}). This trade-off, however, may be worthwhile in view of the numerous other advantageous capabilities of leaky-wave structures including sensing \cite{Ghasempour2020} and frequency multiplexing \cite{Karl2015}.

\subsubsection{Horn Antenna As A Counter Example} \label{Note_3}


\begin{figure}[ht]
  \centering \subfloat[] {\includegraphics[trim={0cm 0cm 0.5cm 0cm},clip, width=0.4\textwidth]{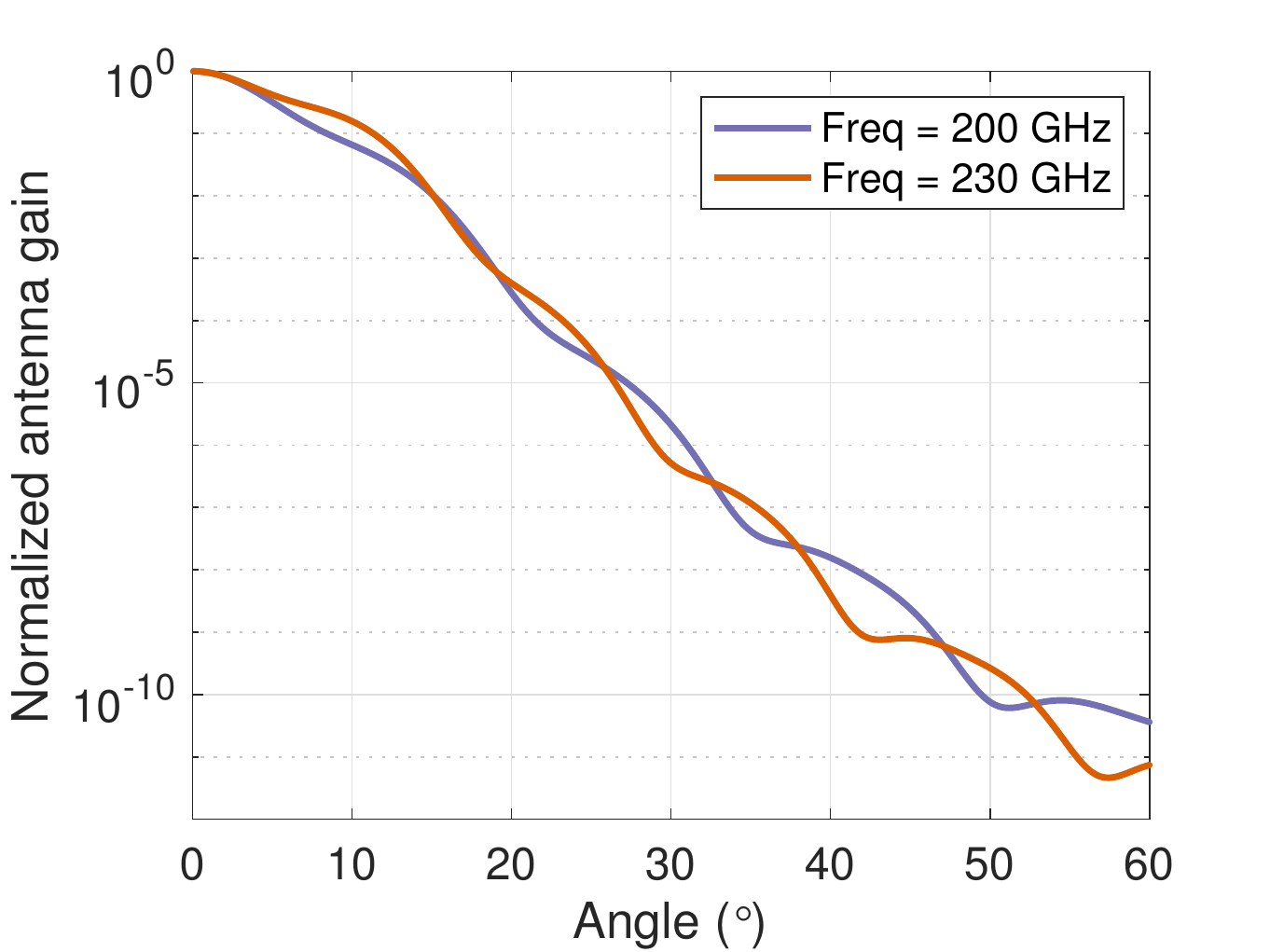}\label{fig:HornRadPattern}}
  \quad
  \subfloat[] {\includegraphics[trim={0cm 0cm 0.5cm 0cm},clip, width=0.4\textwidth]{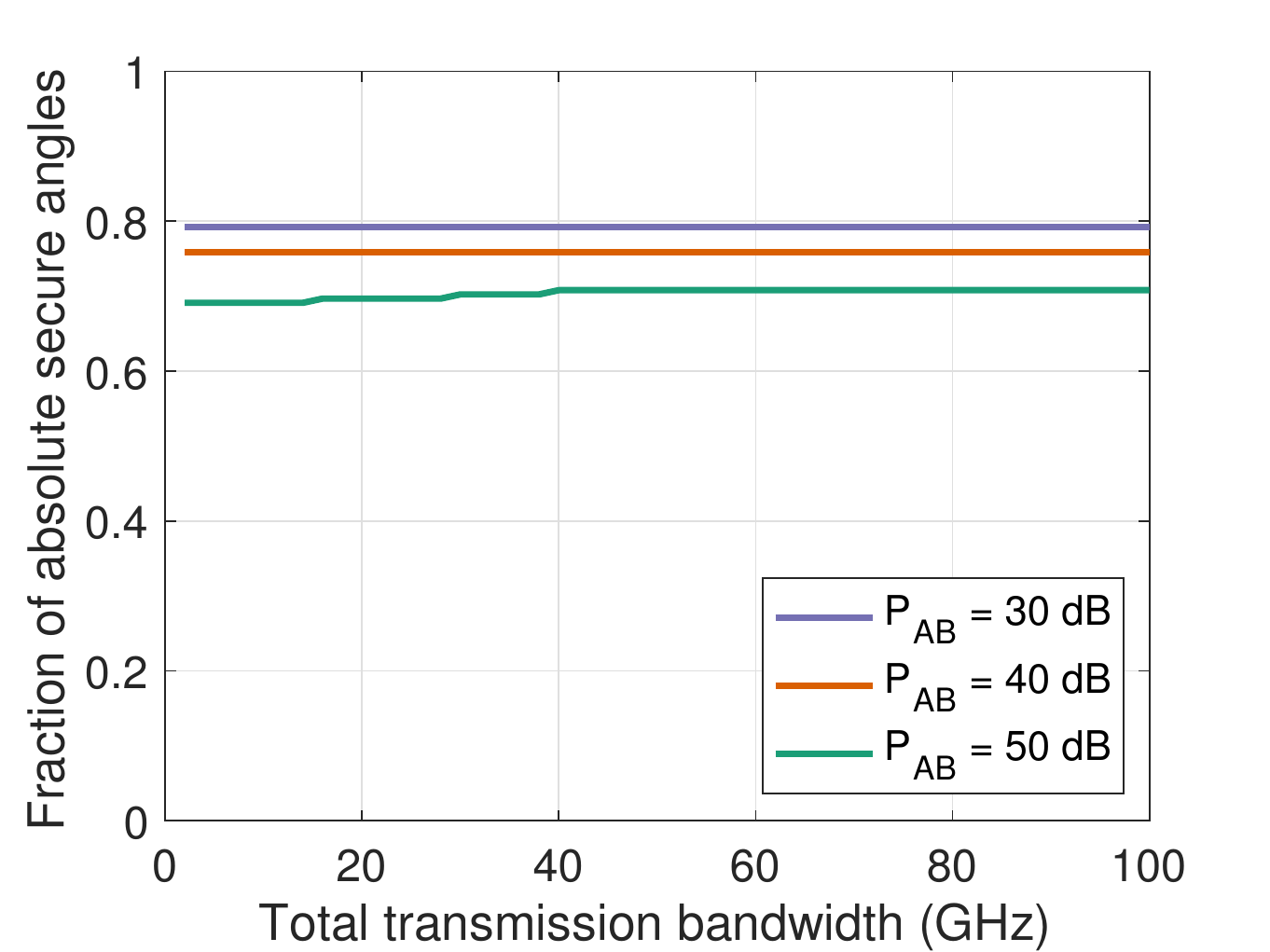}\label{fig:HornSecAng}}
  \caption{Horn antenna. (a) The H-plane radiation pattern from a diagonal horn antenna, computed at two different frequencies. Unlike the antenna patterns shown in Fig. \ref{fig:RadPattn_Para_PA}, this pattern exhibits no pronounced minima. (b) The fraction of secure angles, similar to Fig. \ref{fig:SecAng_TotalBW}, for the horn antenna. Since there are no pronounced minima, there is no improvement with increasing spectral bandwidth. As a result, the creation of blind regions is ineffective, if this antenna is employed without modification.}
\label{fig:HornAntenna}
\end{figure}

The schemes for implementing secure communications in the case where Eve is in the blind region ($\Gamma > 0$) rely on features of the radiation patterns inherent to the antenna used by Alice, specifically the fact that, in certain broadcast directions, these patterns exhibit pronounced minima (or even analytic zeros), due to destructive interference. It is important to realize that this is not a feature of all antennas. Here, we present a counterexample to illustrate this point: a diagonal horn antenna, another commonly employed design in millimeter-wave and terahertz systems.

As in the cases discussed above, the radiation pattern from this antenna, at a given frequency, is also amenable to direct calculation \cite{Johansson1992}. In the calculation, we employ a diagonal horn with a horn length of 20 mm and a diagonal aperture of 11 mm.
Fig. \ref{fig:HornRadPattern} shows one such calculation, in which it is quite clear that the `minima' between any two side lobes (or between the main lobe and first side lobes) are not very pronounced. Fig. \ref{fig:HornSecAng} shows a blind region calculation analogous to the one shown in Fig. \ref{fig:SecAng_TotalBW}, for this horn antenna. This result demonstrates that the blind region does not grow with increasing transmission bandwidth. Thus, the selection of antenna configuration is a key aspect of implementing the proposed security protocol for the blind region.

\subsection{Experimental demonstration} \label{Experimental}

As noted, achieving absolute security requires that the broadcast antenna exhibit pronounced minima whose angular positions vary as a function of frequency. To illustrate the ease with which this can be accomplished, we assemble a link test bed using a horn antenna as the transmitter.
Despite the lack of pronounced minima of horn antennas as observed in Fig. \ref{fig:HornRadPattern}, it is still possible to demonstrate the feasibility of the absolute security system using a horn antenna.

\begin{figure}[t]
  \centering \includegraphics[trim={4.0cm 4.0cm 11cm 4.5cm},clip, width=0.48\textwidth]{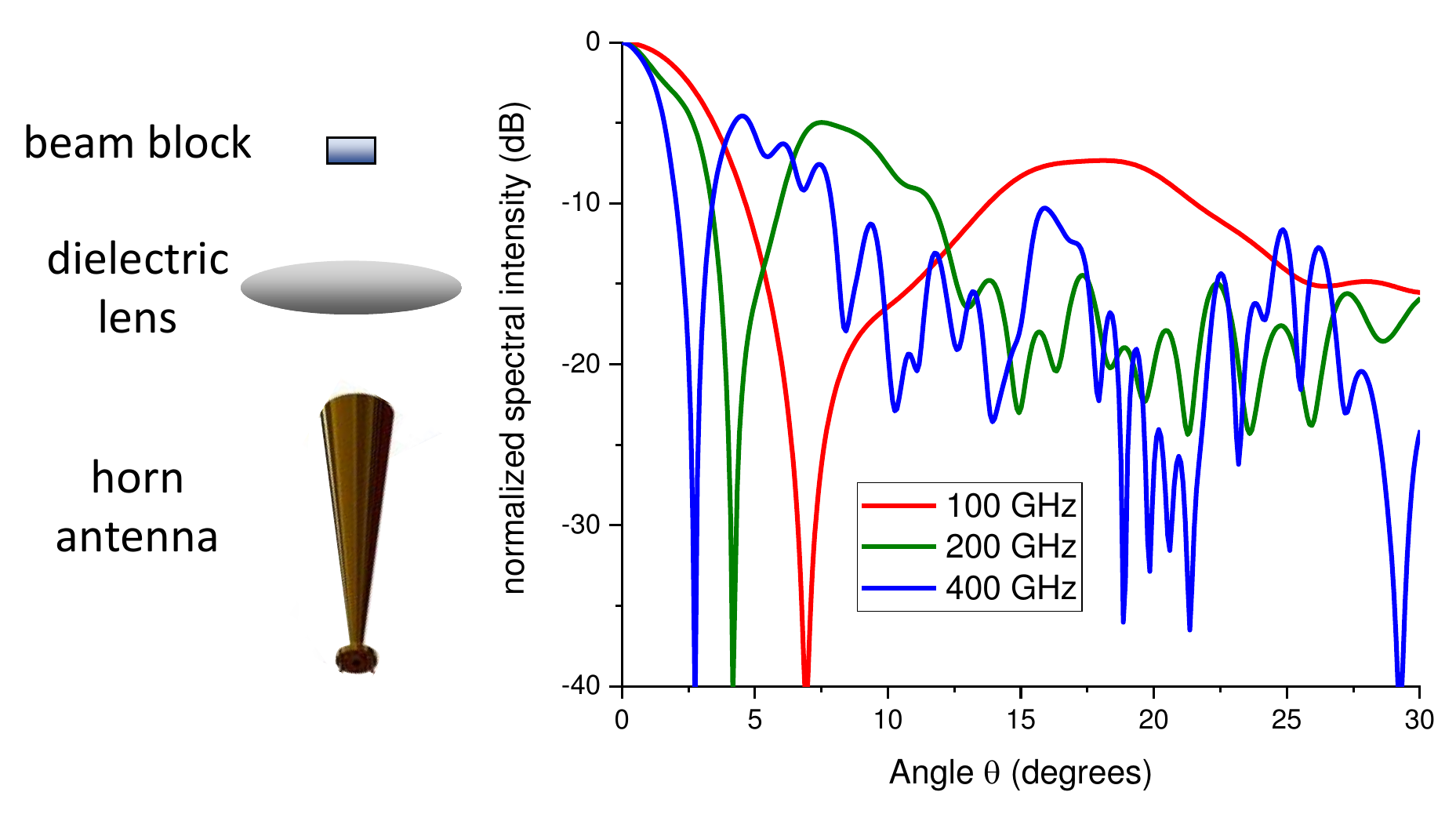}
  \caption{A schematic of the experimental setup used in the measurements described in the above text, and also used in the simulations shown here. These are finite-element simulations of the angular dependence of the far-field diffraction pattern produced by a horn antenna focused on a 4-mm-wide metal beam block. The simulations are performed at the three different frequencies (100, 200, and 400 GHz) used in the experiments.
}
\label{fig:HornWithBlock}
\end{figure}

\begin{figure*}[ht]
  \centering \includegraphics[trim={2.0cm 4.5cm 2.3cm 4.5cm},clip, width=0.85\textwidth]{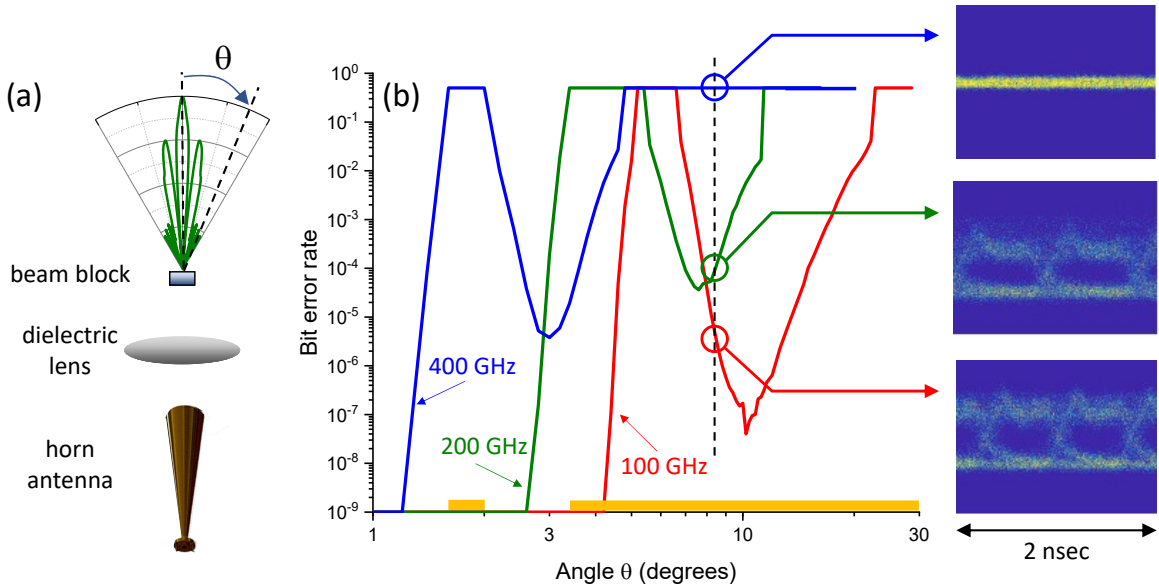}
  \caption{\textbf{Experimental realization of absolute security.} (a) A schematic of the experimental setup. The emission from a horn antenna is focused onto a 4-mm wide beam block, to produce a far-field radiation pattern exhibiting a pronounced minimum at an angle which depends on frequency. The pattern at 200 GHz, computed using a finite element solver, is shown. 
  (b) At three widely spaced frequencies (100, 200, and 400 GHz), a data stream (modulated with on-off keying, at a rate of 1 Gb/sec) is broadcast from the emitter horn antenna, and the bit error rate is measured vs. angle. 
  Eye diagrams for the three frequencies are shown for a representative angle of $\theta = 8.5^{\circ}$ where an eavesdropper could be located. This configuration, using only three channels, creates blind regions for $1.6^{\circ} < \theta < 2.0^{\circ}$ and $\theta > 3.4^{\circ}$ (indicated by the orange bars along the horizontal axis).
}
\label{fig:BER}
\end{figure*}

As illustrated by the schematic in Fig. \ref{fig:HornWithBlock}, we can place a focusing optic (a dielectric lens) in front of the horn, and focus its output onto a diffracting object, in this case a 4-mm-wide metal beam block. The far-field diffraction pattern from this illuminated beam block exhibits a strong maximum on the optic axis (the main lobe, at $\theta = 0$) and a pronounced minimum due to destructive interference at a non-zero angle. Using finite-element simulations, Fig. \ref{fig:HornWithBlock} illustrates the far-field pattern of the setup at three frequencies, 100, 200, and 400 GHz.
Fig. \ref{fig:HornWithBlock} clearly shows the pronounced minimum at a small angle, followed by a subsidiary maximum (first side lobe) at a larger angle.
We note that the first side lobes all peak within 10 dB of the main lobe. Thus, an eavesdropper outside of the main lobe is easily able to detect signals in the individual side lobes, but cannot decode any information from signals at the angles of the minima.
Because these three minima do not coincide with each other, they collectively are expected to form a substantial (though not complete) blind region for angles outside of the main lobe.

To demonstrate the blind region, we perform the experiments employing a frequency multiplier chain in order to generate modulated signals (on-off keying at 1 Gb/sec) at the three widely spaced frequencies (100, 200, and 400 GHz). The modulated data stream is broadcast from the emitter horn antenna, and the bit error rate is measured vs. angle. Fig. \ref{fig:BER} illustrates the experimental arrangement, and shows the measured bit error rates (BER) as a function of angle for a broadcast employing three frequency channels.

At $\theta = 0^{\circ}$ (Bob’s location), we find BER $< 10^{-9}$ at all three frequencies. As $\theta$ increases, each frequency band passes through the minimum of the radiation pattern, where the BER increases to 0.5 (i.e., it is impossible to tell the difference between a ‘0’ and a ‘1’). As $\theta$ increases further, the first side lobe maximum is reached, and the BER again falls to a relatively low value, before once again increasing as the angle increases beyond the edge of the diffracted beam pattern. Eye diagrams for the three frequencies are shown for a representative angle of $\theta = 8.5^{\circ}$ where an eavesdropper could be located. The eye diagrams unambiguously demonstrate that an eavesdropper at this location receives information in only two of the three bands.

Based on the experiments, the blind regions (i.e., the angular locations where at least one frequency is below detection) are indicated by the orange bars along the horizontal axis in Fig. \ref{fig:BER}.
This configuration, using only three channels, creates blind regions for $1.6^{\circ} < \theta < 2.0^{\circ}$ and $\theta > 3.4^{\circ}$.
Even though only three channels are employed, we nevertheless induce a substantial (though not complete) blind region.

%% file: conclusions.tex
\section{Conclusions}

\off{Although we have shown that the blind region can encompass a significant portion of Eve's possible locations, it is clear that it cannot cover them all. However, one could combine our method with the ideas from \cite{cohen2021network} to enable security in the non-blind region as well, while still preserving the favorable scaling with increasing transmission bandwidth.}

It is worth noting that our approach requires engineering of both the physical properties of the transmission system and the data encoding scheme. It is therefore neither purely cryptographic nor purely a physical-layer security system. The hybrid nature of this concept is, to the best of our knowledge, unique in wireless system architectures. We also note that the security guarantees described here are relatively straightforward to achieve, relying only on the assertion that Eve's ability to receive signals is limited by the thermal radiation from the scene she is observing, implying that there exists a smallest measurable signal threshold $\delta > 0$. Apart from that assertion, our analysis affords Eve every strength, such as quantum computing and quantum-noise limited detection. This is the first example of a security protocol which exploits aspects of the physical layer but does not rely on any assumption about noise. We also note that the encoding scheme used by Alice can be known to all, including Eve, without changing any of our conclusions. While our analysis here is based on an idealized propagation environment (free space) and simplified Eve (non-mobile, non-colluding, negligible aperture, etc.), our methods can be extended and optimized for more general scenarios encompassing richer models.